\pdfoutput=1

\def\compileforpublish{1}
\def\isaccepted{1}

\documentclass[conference]{IEEEtran}
\IEEEoverridecommandlockouts

\usepackage{tabularx,calc}
\usepackage{color}
\usepackage{tubscolors}
\usepackage{adjustbox}
\usepackage{tikz}
\usetikzlibrary{%
	shapes,
	automata,
	arrows,
	matrix,
	backgrounds,
	fit,
	patterns,
	decorations.markings, 
	svg.path, 
	shapes.multipart, 
	external, 
	shadows, 
	positioning, 
	calc, 
	decorations, 
	intersections, 
	angles, 
	quotes,
	decorations.pathmorphing, 
	arrows.meta, 
	patterns, 
	decorations.pathreplacing, 
	external,
	snakes%
}

\usepackage{tkz-euclide}
\tikzexternaldisable 

\usepackage{pgfplots}
\usepackage{pgfplotstable}
\usepgfplotslibrary{groupplots}

\usepackage{tikz-3dplot}

\usepackage{siunitx}
\usepackage[
			bibstyle=ieee,
			citestyle=numeric-comp,
			backend=bibtex,
			minnames=1,
			maxcitenames=2,
			maxbibnames=99,
			doi=true,
			isbn=false,
			url=false,
			natbib=true]
			{biblatex} 
\usepackage[caption=false, font=footnotesize]{subfig}
\usepackage{hyperref}
\usepackage{ifthen}
\usepackage{lipsum}
\usepackage{xstring}
\usepackage[abspath]{currfile}
\usepackage{bm} 		
\usepackage{amssymb} 	
\usepackage{amsmath}
\usepackage{siunitx}
\usepackage{graphicx}
\usepackage{bm} 		
\usepackage{booktabs} 	
\usepackage{multirow}   
\usepackage{booktabs} 	
\usepackage{colortbl}
\usepackage{array}
\usepackage{microtype}
\usepackage{here}
\usepackage{tabularx}
\ifx \isaccepted \undefined
\newcommand\copyrighttext{%
	\footnotesize \centering This work has been submitted to the IEEE for possible publication.\\ Copyright may be transferred without notice, after which this version may no longer be accessible.}
\else
\newcommand\copyrighttext{%
	\footnotesize \parbox[t]{.11\textwidth}{\copyright{} \the\year~IEEE.} \parbox[t]{.89\textwidth}{Personal use of this material is permitted. Permission from IEEE must be obtained for all other uses, in any current or future media, including reprinting/republishing this material for advertising or promotional purposes, creating new collective works, for resale or redistribution to servers or lists, or reuse of any copyrighted component of this work in other works.}}
\fi
\newcommand\copyrightnotice{%
	\ifx \compileforpublish \undefined
	\else
	\begin{tikzpicture}[remember picture,overlay]
	\node[anchor=south,yshift=10.5pt] at (current page.south) {\parbox{\dimexpr\textwidth-\fboxsep-\fboxrule\relax}{\copyrighttext}};
	\end{tikzpicture}%
	\fi
}
\newcommand{\odd}{q}
\newcommand{\pmm}{\mathcal{T}}
\newcommand{\para}{\mathcal{P}}

\newcommand{\MATLAB}{\textsc{M\footnotesize ATLAB\normalfont}}
\newcommand{\CASADI}{\textsc{C\footnotesize ASADI\normalfont}}

\newcommand{\MATLABSIMULINK}{\textsc{M\footnotesize ATLAB\normalfont }/\textsc{S\footnotesize IMULINK\normalfont}}

\newcommand\Tstrut{\rule{0pt}{2.6ex}}         
\newcommand\Bstrut{\rule[-1.35ex]{0pt}{0pt}}   

\renewcommand{\vec}[1]{\mathbf{#1}} 
\newcommand{\mat}[1]{\mathbf{#1}} 

\newcommand{\vxv}{v_x^V}

\newcommand{\vyv}{v_y^V}

\newcommand{\psidot}{\dot{\psi}}

\newcommand{\m}{\,\mathrm{m}}
\newcommand{\oom}{\,\frac{1}{\mathrm{m}}}

\newcommand{\mss}{\,\frac{\mathrm{m}}{\mathrm{s}^2}}
\newcommand{\tmss}{\,\mathrm{m} / \mathrm{s}^2}
\newcommand{\kmh}{\frac{\mathrm{km}}{\mathrm{h}}}
\newcommand{\tkmh}{\,\mathrm{km} / \mathrm{h}}

\usepackage[english]{babel}

\addto\captionsenglish{}
\addto\captionsenglish{}
\def\BibTeX{{\rm B\kern-.05em{\sc i\kern-.025em b}\kern-.08em
    T\kern-.1667em\lower.7ex\hbox{E}\kern-.125emX}}
\begin{document}

	\title{ODD-Centric Contextual Sensitivity Analysis Applied To A Non-Linear Vehicle Dynamics Model}

	\author{\IEEEauthorblockN{Richard Schubert}
	\IEEEauthorblockA{\textit{Institute of Control Engineering} \\
	\textit{TU Braunschweig}\\
	Braunschweig, Germany \\
	schubert@ifr.ing.tu-bs.de}
	\and
	\IEEEauthorblockN{Marcus Nolte}
	\IEEEauthorblockA{\textit{Institute of Control Eng.} \\ 
	\textit{TU Braunschweig}\\
	Braunschweig, Germany \\
	nolte@ifr.ing.tu-bs.de}
	\and
	\IEEEauthorblockN{Arnaud de La Fortelle}
	\IEEEauthorblockA{\textit{Centre for Robotics} \\
	\textit{Mines Paris -- PSL}\\
	Paris, France \\
	arnaud.de\_la\_fortelle \\ @minesparis.psl.eu}
	\and
	\IEEEauthorblockN{Markus Maurer}
	\IEEEauthorblockA{\textit{Institute of Control Engineering} \\ 
	\textit{TU Braunschweig}\\
	Braunschweig, Germany \\
	maurer@ifr.ing.tu-bs.de}
	}

	\maketitle
	
	\begin{abstract}%
		Advanced driving functions, for assistance or full automation, require strong guarantees to be deployed.
This means that such functions may not be available all the time,
like now commercially available SAE Level 3 \parencite{sae_j3016_2021} modes that are made available only on some roads and at law speeds.
The specification of such restriction is described technically in the Operational Design Domain (ODD)
which is a fundamental concept for the design of automated driving systems (ADS).
In this work, we focus on the example of trajectory planning and control
which are crucial functions for SAE level 4+ vehicles and often rely on model-based methods.
Hence, the quality of the underlying models has to be evaluated with respect to the ODD.
Mathematical analyses such as uncertainty and sensitivity analysis support the quantitative assessment of model quality in general.
In this paper, we present a new approach to assess the quality of vehicle dynamics models using an ODD-centric sensitivity analysis.
The sensitivity analysis framework is implemented for a 10-DoF nonlinear double-track vehicle dynamics model
used inside a model-predictive trajectory controller.
The model sensitivity is evaluated with respect to given ODD and maneuver parameters.
Based on the results, ODD-compliant behavior generation strategies with the goal of minimizing model sensitivity are outlined.
	\end{abstract}%
	
	\copyrightnotice
	
	\section{Introduction}
\label{sec:intro}
The ODD is a key concept in the design process of an ADS
and specifies the operating conditions under which the ADS is designed to function \parencite{sae_j3016_2021},
e.g. including possible road types, weather conditions and traffic participants.
Following ISO 21448 (SOTIF) \parencite{international_organization_for_standardization_iso_2022},
the functional boundaries of a system and each of its subsystems must be evaluated \parencite{colwell_automated_2018}.
The system must respond appropriately if ODD boundaries are violated at runtime,
e.g. in terms of an undesired change of road or weather conditions.
In this context, self-awareness is a powerful concept which, e.g., incorporates system health and ODD monitoring.
To achieve self-awareness, knowledge of the system's architecture, capabilities and dynamics must be stored in models \parencite{nolte_supporting_2020}.
These models must be valid under the operating conditions of the ADS to be able to pursue the mission goal of the automated vehicle's operation \parencite{gregory_self-aware_2016}.
For an ODD to be properly defined and validated, models have to be efficient, but additionally their validity has to be monitored and assessed as a part of the system’s self-awareness.

In this paper, we consider the example of vehicle dynamics models which are often used in motion planning and control.
Optimal trajectories with respect to vehicle dynamics can be planned and controlled using the model knowledge
(e.g. \parencite{nolte_model_2017, stolte_reference_2019, stolte_towards_2023}).
The quality of such approaches directly depends on the quality of the underlying vehicle models.
Insufficient model quality does not only reduce the quality of the system's performance
but can even lead to the violation of safety constraints which may be satisfied by the vehicle model,
e.g. if they are explicitly considered in model-based motion planning and control algorithms,
but could be violated by the real vehicle from whose actual dynamics the model deviates.
As many influences on vehicle dynamics can be related to properties of the ODD, it is crucial to ensure the model's validity inside the chosen ODD.
The model quality can degrade with respect to the applied dynamics \parencite{polack_kinematic_2017}
which themselves are related to the operating conditions of the ADS
since the specified ODD has an impact on the model's parameter space as well as on the inputs to which the model is exposed.
Therefore, it must be ensured that invalid models do not cause the automated vehicle to violate ODD boundaries.

In mathematics, general methods for the assessment of model quality exist,
such as sensitivity and uncertainty analysis \parencite{dickinson_sensitivity_1976, loucks_water_2017}.
They allow to quantify the sensitivity of the model with respect to its parameters
and evaluate the impact of parameter uncertainty on model behavior.
In this context, the sensitivity of the model with respect to its parameters and the resulting potential
to propagate uncertainties are used as a measure for model quality.
For example, a low sensitivity of the model implies that measurement errors
do not have a severe impact.
Therefore, sensitivity analysis is a method used in the design process, e.g.,
to identify sensitive parameters which have a great influence on the model's behavior.
The quantitative results of the analysis potentially
allow to even derive requirements for the measurement accuracy of the model parameters.
Moreover, the analysis can help to find a valid model
that can be applied in the specified ODD with sufficiently low sensitivity.

In this paper, based on previous work \parencite{nolte_sensitivity_2020},
we provide the framework for an ODD-centric contextual sensitivity analysis for vehicle dynamics models.
Our approach demonstrates how the sensitivity analysis can be deployed within a specified ODD with the constraints of a
given driving maneuver and how it helps to assess a model-based trajectory controller.
As an outlook,
we consider our findings from a runtime perspective and propose to minimize the sensitivity inside the controller
through the constraints of the applied dynamics at runtime.
Therefore, implications on sensitivity-oriented runtime constraints for a self-aware
model-based planning and control architecture
as well as on the higher-level behavior generation are outlined in this paper.

The paper is structured as follows:
In \autoref{sec:literature}, we present previous work on which this work has been built and the progress we have made over the state of the art.
In \autoref{sec:odd}, we state more precisely what the ODD is.
We choose a subset of ODD elements that have a high impact on vehicle dynamics and that will represent the ODD in the remainder of this paper.
Within this ODD, several types of maneuvers can take place. We select one for the purpose of the analysis: a lane change.
Based on a derived set of ODD and maneuver parameters, we present the sensitivity analysis framework in \autoref{sec:application}.
In a simulation, reference trajectories are generated and fed into the model-predictive trajectory controller.
The sensitivity analysis is performed based on the resulting model states and results are presented in \autoref{sec:results}.
The implications of our approach are concluded in \autoref{sec:conclusion} and an outlook is given.

\begin{figure*}[t]
    \centering
	\includegraphics[width=0.94\textwidth]{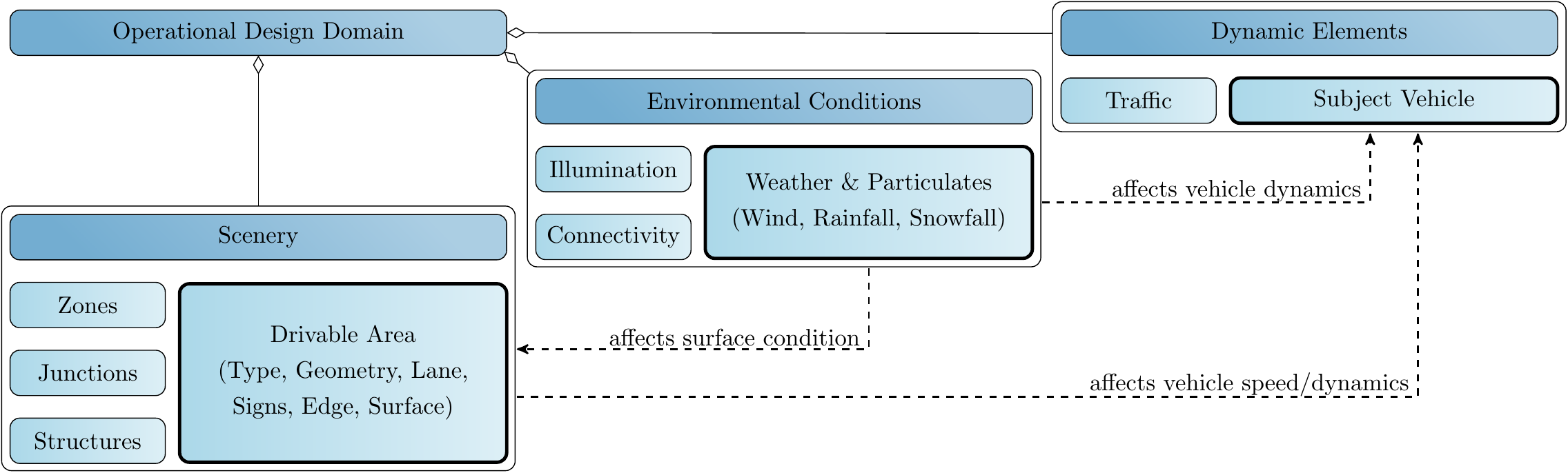}
    \caption{Selected ODD elements as defined in \parencite{bsi_operational_2020} and their impact on vehicle dynamics}
    \label{fig:odd}
\end{figure*}
	\section{Related Work} \label{sec:literature}
As also noted in previous work \parencite{nolte_sensitivity_2020}, sensitivity and uncertainty analysis are two mathematical methods
that allow the quantitative assessment of model quality.
Several publications in the field of model theory and control engineering focus solely on sensitivity analysis,
even though \textcite[p. 331-334]{loucks_water_2017} introduce sensitivity and uncertainty analysis as two methods
that must be combined to gain full insights into a model's quality.
\textcite{reuter_how_2011} propose to assess model quality
through the evaluation and documentation of a sensitivity analysis.
The formal definition of a direct method for local sensitivity analysis
with respect to the parameters of a dynamic system is given by \textcite{dickinson_sensitivity_1976}.
Other publications focus on the explicit consideration of model sensitivity at a functional level:
In the work of \textcite{yedavalli_controller_1982},
the quantified sensitivity is introduced into the cost function of a model-based controller
to minimize model sensitivity in the control loop.
\textcite{tulpule_integrated_2014} uses their work to develop a framework for
the design of robust controllers through the integration of sensitivity minimization.
Their work joins other contributions that aim to compensate model errors in the controller design
(e.g. \parencite{simkoff_plant-model_2017, thangavel_handling_2018, polack_guaranteeing_2018}).

While most papers present general methods in the intersection of control engineering and model theory,
only a few publications explicitly refer to the application of a sensitivity analysis in the context of vehicle dynamics models:
In the work of \textcite{jang_state_1997} and \textcite{hamza_contribution_2015}, a sensitivity analysis is applied to a single-track model.
\textcite{nolte_sensitivity_2020} perform a sensitivity analysis for a linear single-track and a nonlinear double-track model.
In contrast to previous literature, realistic trajectories from an urban ODD are considered for the analysis and connections to the field of automated driving are drawn.
\textcite{nolte_sensitivity_2020} show that the sensitivities of the two models do not deviate greatly which,
despite the reduction of model complexity, does not imply much stricter requirements on parameter accuracy for the linear model.
Furthermore, their analysis is evaluated in case of an exemplary actuator degradation.
However, the sensitivity analysis in \parencite{nolte_sensitivity_2020} is evaluated for a large set of trajectories
without explicitly considering the ODD or the dynamic constraints applied in the generation process.
Furthermore, the quantitative results of the sensitivity analysis are evaluated in a rather coarse fashion
since the large set of resulting sensitivity signals cannot be reviewed collectively to identify relevant characteristics,
i.e. no aggregated measure is introduced that would allow to summarize the information.

Therefore, in this paper, the results of previous work \parencite{nolte_sensitivity_2020} are extended and further structured:
The scope of this paper is the application of an ODD-centric sensitivity analysis for
a nonlinear 10-DoF double-track vehicle dynamics model that is used in a trajectory controller.
Through the consideration of specific ODD and maneuver parameters, context is added to the analysis.
By calculating and aggregating the model sensitivity, the results can be directly related to these parameters.
	\section{ODD and Maneuvers}
\label{sec:odd}
Any ADS is developed with a specific ODD in mind which acts as a foundation of the design process.
According to SAE J3016, the term ODD is defined as the
``operating conditions under which a given [ADS] or feature thereof is specifically designed to function,
including, but not limited to, environmental, geographical, and time-of-day restrictions, and/or the requisite presence or
absence of certain traffic or roadway characteristics'' \parencite[p. 17]{sae_j3016_2021}.
As stated in the introduction of this paper,
any system component or function must be suitable for being used inside the defined ODD,
including model-based motion planning and control algorithms \parencite{stolte_reference_2019}.
In the following, we therefore examine how the dynamics subjected to the model and its parameters
relate to the ODD elements.

\subsection{Operational Design Domain}\label{sec:odd_sub}
According to \parencite{bsi_operational_2020} and consistent with the 5- \parencite{bagschik_ontology_2018} and 6-layer models
\parencite{scholtes_6-layer_2021} for scenario description,
the defined ODD can be divided into the following parts: \emph{scenery}, \emph{environmental conditions} and \emph{dynamic elements}.
The scenery specifies the drivable area's type, geometry, surface condition as well as lane markings and road signs.
Different driving zones, e.g. geo-fenced areas or traffic management zones, junctions
as well as temporary, fixed and special structures are included.
Environmental conditions summarize weather conditions, e.g. wind, rainfall and snowfall, and particulates such as sand and dust.
Also, illumination, e.g. due to daytime, and connectivity such as V2X-communication
and positioning services are considered.
Dynamic elements cover the presence of traffic, e.g. its density, flow rate and the types of participants.
Also, the ego-vehicle itself is defined as a dynamic element.

With respect to the ODD definition in \cite{bsi_operational_2020},
a subset of ODD elements with high relevance for vehicle dynamics is shown in \autoref{fig:odd}:
The drivable area's geometry, e.g. curvature, camber and slope, and surface directly influence the vehicle's dynamics.
The drivable area's type may be associated with a speed limit which determines the vehicle's speed range.
Weather conditions and particulates affect the condition of the drivable area's surface and hence the road-tire friction.
Wind represents an additional force directly affecting the vehicle's dynamics.
In terms of vehicle dynamics models, these elements are either incorporated into the model's parameter space or the applied dynamics.
If model-based vehicle control algorithms are used, target speed and speed constraints
set in the controller must lay within the speed range defined in the ODD.
Another example is the road-tire friction quantified through the road-tire friction coefficient:
On the one hand, it is an ODD attribute describing the drivable area's surface condition and,
on the other hand, it is a parameter of the vehicle dynamics model.

\subsection{Maneuvers}\label{sec:beh_sub}
\begin{figure}[!h]
    \centering
    \vspace{-0cm}
    \includegraphics[trim = 0 0 0 2em, clip, width=0.96\columnwidth]{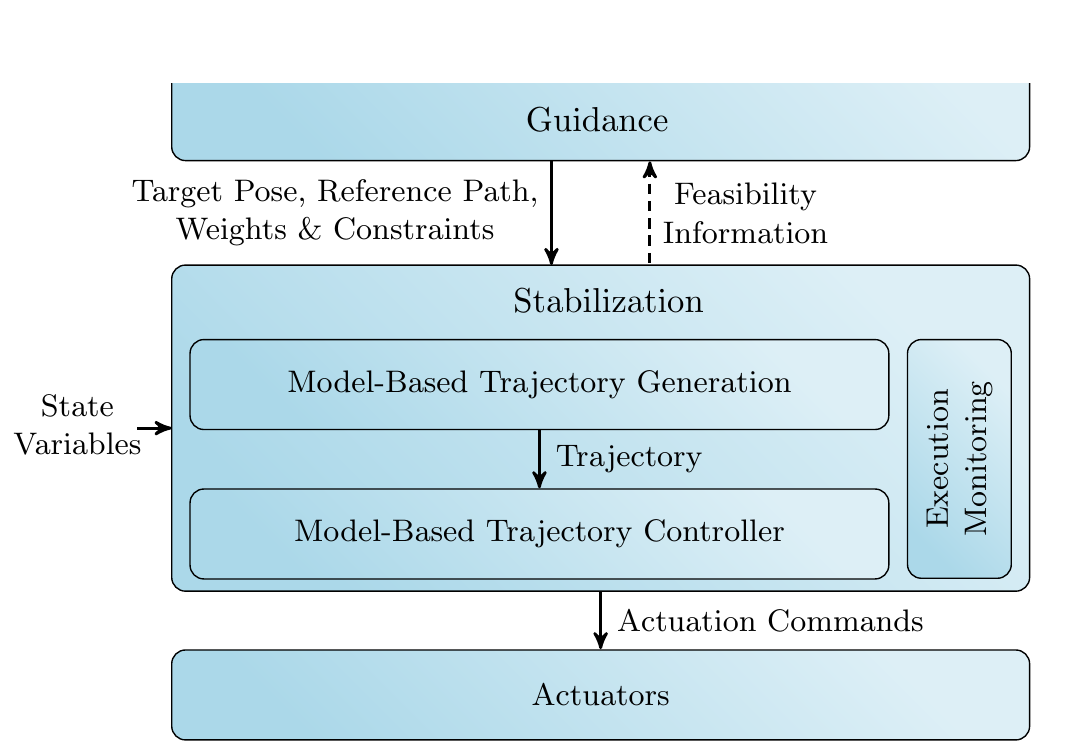}
    \vspace{-0.3cm}
    \caption{Guidance and stabilization layer according to
        \parencite{nolte_sensitivity_2020}, based on \parencite{nolte_model_2017,ulbrich_towards_2017}}
    \label{fig:architecture}
\end{figure}

Within the defined ODD, the automated vehicle can make different tactical decisions and thus move in different ways.
Even though the behavior of the vehicle depends on the ODD, i.e. in terms of the abovementioned conditions for the vehicle's dynamics,
a range of different movements depending on the automated vehicle's tactical decision can be performed.
In other words, the driving dynamics of the vehicle mainly depend on its behavior but are constrained through the defined ODD.
Therefore, the vehicle's tactical behavior should be considered in more detail as well.
Following their remarks on the definition of an ODD,
\textcite{koopman_how_2019} introduce maneuvers to structure the automated vehicle's external behavior
and consider them to be necessary for the complete validation of an ADS.
In particular, according to \textcite{jatzkowski_zum_2021}, a \emph{driving maneuver}
is defined as an abstraction of possible state progressions for the motion of a vehicle.
A particular sequence of dynamic states represents a particular movement of the vehicle, i.e. a specific \emph{instance} of the maneuver.
In a functional architecture as described by \parencite{ulbrich_towards_2017} and partly visualized in \autoref{fig:architecture},
an appropriate driving maneuver is selected in response to the current situation at the guidance layer.
The output of the maneuver information is extended by certain constraints, e.g. target pose, reference path and additional weights.
This instantiation of the maneuver is performed through the generation of a target trajectory at the stabilization layer
while respects derived constraints.
By applying the target trajectory to a trajectory controller, the desired motion can be realized.

	\section{Sensitivity Analysis}
\label{sec:application}
Sensitivity analysis is a mathematical tool which allows to determine the impact of the model parameters on the model's behavior quantitatively.
Assuming measurement imperfection and hence inaccurate parameter values in general,
sensitivity can be seen as the model's potential to propagate parameter errors into the model behavior.
Hence, when applying model-based methods, areas of high sensitivity should be avoided if high accuracy is required.
In this paper, we focus on analyzing a model-based trajectory controller as part of
a model-based motion planning and control architecture that is located at the stabilization layer in \autoref{fig:architecture}.
For example, the required inputs to follow a target trajectory can be computed by using a model-predictive control (MPC) algorithm
which makes model-based predictions about the vehicle's dynamics with respect to possible model inputs.
The state of the real vehicle is then compared to the predicted state in the next step and a new prediction is made.
Model errors, e.g. in terms of inaccurately determined model parameters, hence lead to an incorrect prediction.
Even though model errors cannot accumulate along the full trajectory,
the MPC cannot compensate for control errors between the prediction time steps
and suboptimal performance of the controller is expected \parencite{werling_neues_2011, thangavel_handling_2018}.
Due to the complexity of the MPC algorithm and the underlying vehicle dynamics model,
the consequences of a model mismatch cannot be easily estimated.
Thus, safety-critical spatial deviations in the real vehicle's motion
caused by a model mismatch in the controller cannot be ruled out.

\begin{figure}[h]
    \centering
    \includegraphics[width=0.85\columnwidth]{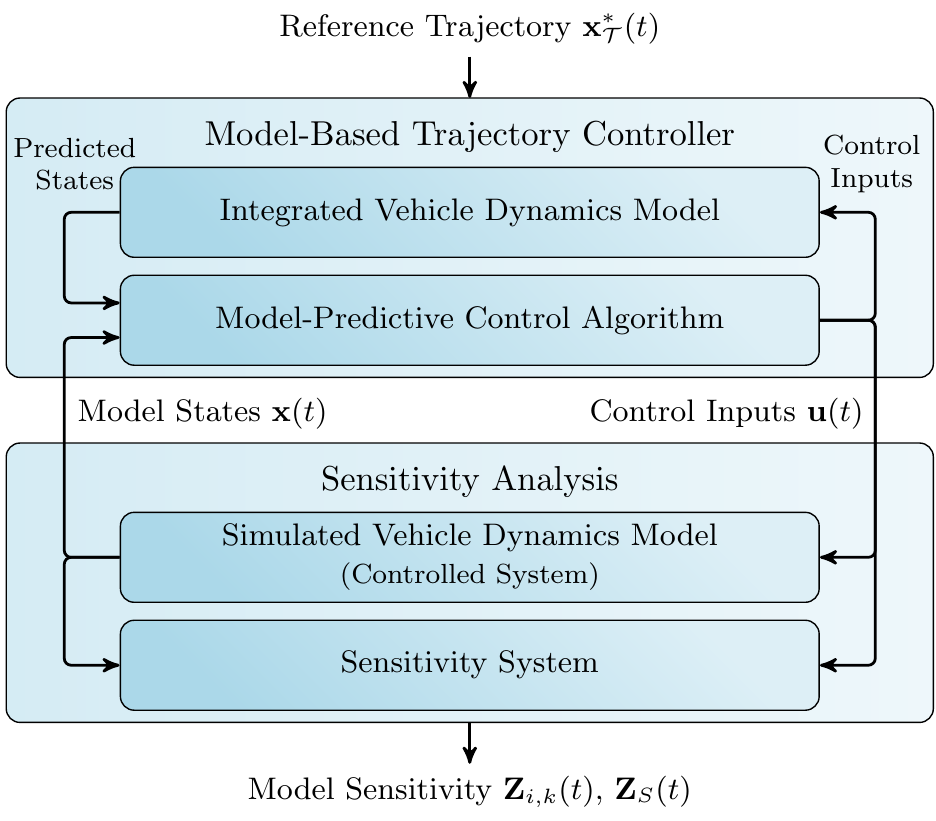}
    \caption{Simulation of the MPC, vehicle dynamics model and sensitivity system}
    \label{fig:setup}
\end{figure}

Following previous work on the consideration of model sensitivity in control design
\parencite{yedavalli_controller_1982, tulpule_integrated_2014},
we aim to develop strategies to avoid areas of high sensitivity in the state space of the vehicle dynamics model inside the controller.
In particular, by using the previously introduced ODD and maneuver definitions,
statements about sensitivity can be related to the properties of the ODD and a selected maneuver.
The goal of the method presented in this paper is to gain a priori knowledge about model sensitivity with respect to various ODD and maneuver parameters.
As stated before, the resulting data can support the design process of model-based methods for motion planning and control.
Furthermore, since the resulting sensitivity can be related to the maneuver attributes,
constraints for the behavior of the automated vehicle with respect to model sensitivity can be derived.

Following the functional architecture shown in \autoref{fig:architecture},
a selected driving maneuver is instantiated through a target trajectory with respect to additional maneuver constraints.
To mimic the interface between the guidance and stabilization layer,
we generate a reference trajectory with respect to selected ODD and maneuver attributes offline in the first step.
Second, the reference trajectory is applied to the model-predictive trajectory controller which generates the required control signals.
The applied MPC algorithm is described by \textcite{stolte_towards_2023}
and utilizes a complex nonlinear double-track vehicle dynamics model.
Note that the chosen vehicle dynamics model and control algorithm act as an example
and we suggest that any other model-based controller could be analyzed using the presented approach.
To determine the sensitivity of the vehicle dynamics model integrated in the MPC algorithm along the reference trajectory,
the trajectory controller is connected to an identical vehicle dynamics model in a closed loop
which mirrors the model integrated in the controller and provides the required feedback.
This simulation setup is shown in \autoref{fig:setup}.
The key idea here is that following the generated reference trajectory induces the same dynamics
in the externally simulated vehicle model as they are present in the integrated model inside the controller.
By providing the controller feedback through an identical model, external dynamics can be excluded.
Moreover, the predicted states of the integrated model over each prediction period can be assumed to be identical to the states of the external model
since both models yield identical dynamics.
This enables us to make statements about the sensitivity of the prediction model
by assessing the states of the externally simulated model along the full trajectory.

\subsection{Selection of ODD and Maneuver Parameters}\label{sec:oddspec}
In order to systematically investigate the sensitivity of the vehicle dynamics model,
we generate reference trajectories with respect to the attributes of the ODD and the vehicle's behavior.
Therefore, we consider our findings from \autoref{sec:odd}
in the selection of an appropriate parameter set.
In the first step, a subset of the previously described ODD attributes is selected,
i.e. only those elements that can be represented in the simulation of vehicle dynamics:
\begin{itemize}
	\item drivable area geometry such as curvature, camber or slope
    \item target speed limit, e.g. determined through road type/signs
    \item friction coefficient representing the road's surface condition (e.g. influenced through weather)
	\item lane width, e.g. as determined through lane markings (equal to the lateral distance between two lane centers)
\end{itemize}
In a second step, we determine an exemplary specification for the considered ODD elements.
The model-based trajectory controller is designed to control the vehicle's horizontal motion under the assumption
that the drivable area's topology is planar inside the specified ODD, i.e. camber and slope are assumed to be zero.
Therefore, we restrict the specified ODD accordingly and solely focus on the vehicle's horizontal motion in the following.
In this paper, the exemplary ODD specification shall cover areas with different dynamic conditions, e.g. inner city and highway sections.
Therefore, we set the maximum speed range of the automated vehicle
to $v_{\odd} \in (0\tkmh,\, 130\tkmh]$ which is the recommended value for German highways.
To cover various environmental and drivable area conditions,
we set the value range for the friction coefficient to \linebreak $\mu_{\odd} \in [0.25,\, 1.1]$ as an example.

In a third step, the instantiation of the maneuver is extended by individual dynamic constraints for the selected maneuver.
We consider the example of a lane change maneuver as defined in \parencite{jatzkowski_zum_2021}.
We further assume that the vehicle is centered in its current lane at the beginning of the maneuver execution
and the target pose is set to the center of an adjacent lane.
As described in \parencite{jatzkowski_zum_2021}, the external behavior is always composed by a longitudinal and a lateral maneuver.
We assume that the lane change (lateral) is here executed in combination with following a desired speed (longitudinal).
Hence, we argue that the resulting maneuver instantiation should at least incorporate the following constraints:
\begin{itemize}
    \item direction of the lane change (left/right)
    \item target speed throughout the maneuver or final speed
    \item maximum acceleration, yaw rate and/or path curvature
\end{itemize}
For example, due to functional limitations of the motion controller,
the selected acceleration limit for a lane change maneuver is restricted to $a_{\odd} \in (0\tmss,\, 6\tmss]$ in this work.
The selected driving maneuver and its parameter set can be understood as
an extended \emph{maneuver template} \parencite{jatzkowski_zum_2021} which in this work includes environmental features.
As a result, the following tuple of ODD and maneuver parameters $m_{\odd}$ is derived for the lane change maneuver:
\begin{equation}\label{eq:modd_lc}
    m_{\odd}\big |_{\mathrm{Lane Change}} = \left ( w_{\odd}, k_{\odd}, \mu_{\odd}, v_{1,\odd}, v_{2,\odd}, a_{\odd} \right )
\end{equation}
with lane width $w_{\odd}$, lane curvature $k_{\odd}$, friction coefficient $\mu_{\odd}$,
the vehicle's initial and final target speed $v_{1,\odd},\, v_{2,\odd}$ as well as the maximum acceleration limit $a_{\odd}$.
We set the ODD and maneuver parameter ranges to the values in \autoref{tab:parameters} for the examples in this paper.
For the sake of brevity, we set $w_{\odd},\, k_{\odd}$ to constant values in all following examples.

\begin{table}[h]
    \centering
    \caption{Considered ODD and maneuver parameters for a lane change}
    \begin{tabular}{ c | c | c | c | c }
        \toprule        
            $w_{\odd}$ & $k_{\odd}$ & $\mu_{\odd}$ & $v_{1,\odd},\; v_{2,\odd}$ & $a_{\odd}$ \Bstrut \\ \hline
            $4\m$ & $0\oom$ & $0.25\,\dots\,1.1$ & $0\kmh\,\dots\,130\kmh$ & $0\mss\,\dots\, 6\mss \Tstrut$ \\
        \bottomrule
    \end{tabular}
    \label{tab:parameters}
\end{table}

\subsection{Reference Trajectory Generation}\label{reference_trajectories}
\begin{figure}[h]
    \centering
    \includegraphics[width=\columnwidth]{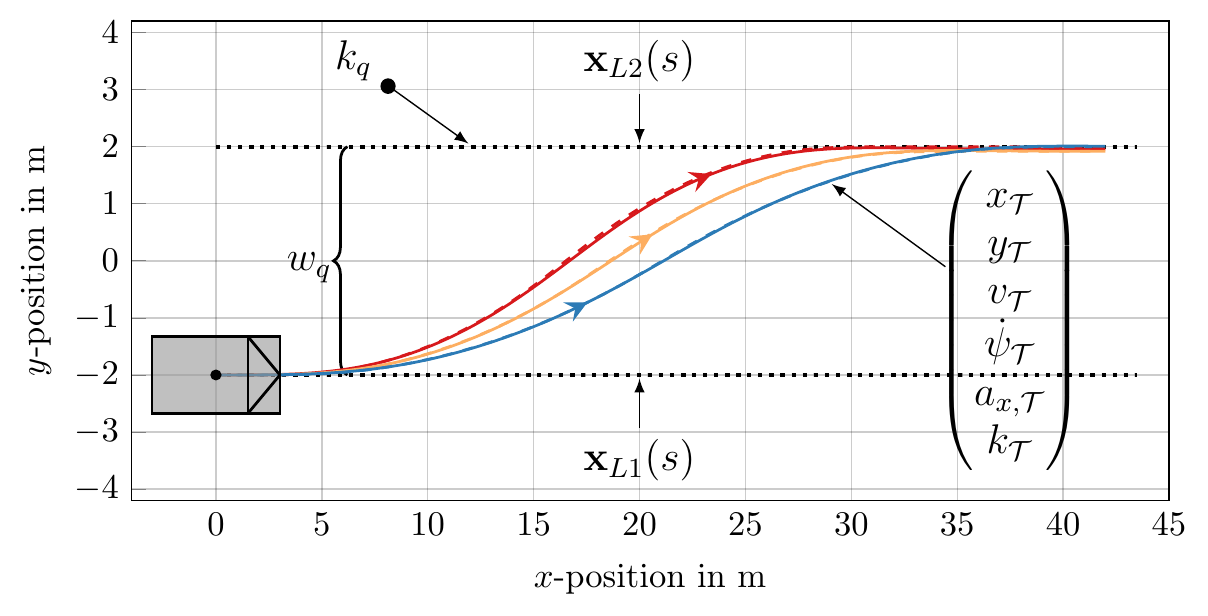}
    \vspace{-0.8cm}
    \caption{Reference (dashed) and simulated path (solid) of the vehicle model
        for a lane change maneuver with acceleration constraint $a_{\odd}=2.5\mss, 3.5\mss, 4.5\mss$}
    \label{fig:reftraj}
\end{figure}

To perform the instantiation of trajectories,
we use an optimization-based offline reference trajectory generator as presented by \textcite{frese_generierung_2019}.
The trajectory generator applied in this work utilizes a simple point mass model and respects derived ODD and maneuver parameters as constraints.
Given these constraints, the optimization algorithm generates a time-minimal trajectory allowing to transform the initial into the target state.
In particular, the problem formulation presented in \parencite{frese_generierung_2019}
uses a point-mass model with the following state equations $\vec{F}_{\pmm} = (\vec{u}_{\pmm}, \vec{x}_{\pmm})$
(time $t$ is neglected for the sake of brevity):
\begin{equation}\label{eq:pmm}
	\vec{\dot{x}}_{\pmm} =
		\vec{F}_{\pmm}\left ( \begin{pmatrix}
			\dot{a}_{x, \pmm} \\
			\dot{k}_{\pmm}
		\end{pmatrix}, \begin{pmatrix}
			x_{\pmm} \\
			y_{\pmm} \\
			v_{\pmm} \\
			\psidot_{\pmm} \\
			a_{x, \pmm} \\
			k_{\pmm}
		\end{pmatrix} \right ) =
		\begin{pmatrix}
		v_{\pmm}\cdot\cos (\psi_{\pmm}) \\
		v_{\pmm}\cdot\sin (\psi_{\pmm}) \\
		a_{x, \pmm} \\
		k_{\pmm}\cdot v_{\pmm} \\
		\dot{a}_{x, \pmm} \\
		\dot{k}_{\pmm}
	\end{pmatrix}
\end{equation}
$x_\pmm$ and $y_\pmm$ describe the position of the point-mass.
$v_\pmm$ is the speed, $\psi_\pmm$ is the yaw angle,
$a_{x, \pmm}$ is the acceleration and $k_\pmm$ is the curvature of the resulting path.
The optimization problem is then defined in \parencite{frese_generierung_2019} as follows:
\begin{equation}\label{eq:pmm_problem}
    \begin{split}
        \min_{\vec{u}_{\pmm},\, s_{\mathrm{End}}} & \quad\;\, T \\
        \mathrm{s.t.}\qquad\qquad \dot{x}_{\pmm}(t) & = \vec{F}(\vec{x}_{\pmm}(t),\, \vec{u}_{\pmm}(t)) \\
        \vec{x}_{L1}(0) & = \vec{x}_{\pmm}(0) \\
        \vec{x}_{L2}(s_{\mathrm{End}}) & = \vec{x}_{\pmm}(T) \\
        0 \leq a_x^2(t)+(k(t)\cdot v^2(t))^2 & \leq a_{\mathrm{max}}^2,\;\forall t \in [0,\, T] \\
        0 \leq v(t) & \leq v_{\mathrm{max}},\;\forall t \in [0,\, T]
    \end{split}
\end{equation}
For the reference trajectory generation, we focus on the aforementioned example of a lane change maneuver.
In this case, the generated trajectory must connect the lane centers of the current and the adjacent lane.
The position of the lane center along a road segment is represented through parabolic curves
which are represented through $\vec{x}_{L1}(s)$ and $\vec{x}_{L2}(s)$ with arc length $s$.
For illustration, possible trajectories with respect to the ODD parameters are visualized
alongside the lane centers (dotted) in \autoref{fig:reftraj}.
In addition to reaching the target position,
the model must comply with the specified initial and target speed $v_{1,\odd},\,v_{2,\odd}$.
The speed and acceleration constraints in \eqref{eq:pmm_problem}
are set based on the maneuver parameters, $v_{\mathrm{max}} = \max (v_{1,\odd},\,v_{2,\odd})$ and $a_{\mathrm{max}} = a_{\odd}$.
The problem formulation in \eqref{eq:pmm_problem} hence covers all parameters in \eqref{eq:modd_lc},
except for the friction coefficient $\mu_{\odd}$ which is directly incorporated in the vehicle dynamics model.
The optimal model inputs $\vec{u}^{*}_{\pmm}(t)$ as well as the arc length of the trajectory $s_{\mathrm{End}}$
are found as the problem's solution in \MATLAB\ using the \CASADI\ library \parencite{andersson_casadi_2019}.
The corresponding time-minimizing sequence of states $\vec{x}^{*}_{\pmm}(t)$ is then stored as the optimal reference trajectory.

\subsection{Simulation of the Vehicle Dynamics Model}\label{simulation}
In a \MATLABSIMULINK\ simulation environment, we apply the generated reference trajectory to the model-based trajectory controller
by \parencite{stolte_towards_2023}
in a closed loop with the nonlinear double-track vehicle dynamics model.
A \textit{Magic Formula Tire Model} \parencite{pacejka_magic_1992}
is integrated in the state-space representation of the vehicle dynamics model in our implementation.
In mathematical terms, the analyzed vehicle dynamics model is a dynamic system
and hence described through a set of state equations $\vec{\dot{x}}(t) = \vec{F}(\vec{u}(t),\vec{x}(t),\para)$
with model states $\vec{x}(t)=(x_1,\dots,x_N)^T$ model inputs $\vec{u}(t)=(u_1,\dots,u_M)^T$, and model parameters $\para\in\{p_1,\dots,p_K\}$.
The full model description yielding $N=10$ states and $K=49$ parameters is given in \parencite{nolte_sensitivity_2020}.
For example, as mentioned before, the friction coefficient $\mu_{\odd}$ is a parameter of the vehicle dynamics model.
In the simulation, the MPC generates a sequence of control inputs $\vec{u}(t)$
for the simulated vehicle dynamics model to follow the reference trajectory.
Based on the resulting model states $\vec{x}(t)$, the model sensitivity is computed as described in the following subsection.
The setup is identical to the one described in \parencite{nolte_sensitivity_2020}.

\subsection{Application of the Sensitivity Analysis}\label{sec:appsens}
In this paper, the sensitivity of a model is quantified as
the impact of a possible parameter variation on the model's behavior.
In mathematical terms,
the sensitivity of state $x_i(t)$ with respect to parameters $p_k$ is hence defined as
the partial derivative of the model state with respect to the parameter \parencite{dickinson_sensitivity_1976},
\begin{equation}\label{eq:sensitivity}
    Z_{i,k}(t)=\frac{\partial x_i(t)}{\partial p_k}.
\end{equation}
To calculate the sensitivity, we follow the definition by \textcite{dickinson_sensitivity_1976}
and incorporate the dynamic system's state-space representation $\dot{x}(t)=\vec{F}(\vec{x}(t),\vec{u}(t),\para)$
in the definition of the sensitivity in \eqref{eq:sensitivity}.
Using a matrix representation for the set of state equations as in \parencite{dickinson_sensitivity_1976},
we solve the sensitivity system
\begin{equation}\label{eq:sensystem}
    \vec{\dot{Z}}_{k}(t)=\vec{F}_{k}(t)+\mat{J}(t)\,\vec{Z}(t)
\end{equation}
numerically.
$\mat{J}(t)(t)$ denotes the Jacobian matrix of the vehicle dynamics model.
$\vec{F}_{k}(t)$ refers to the vector of partial state derivatives with respect to $p_k$,
\begin{equation*}
	\vec{F}_{k}(t)=\nabla_{p_k}\, \vec{F}(\vec{x}(t),\vec{u}(t),\para).
\end{equation*}
The numerical simulation of the sensitivity system results into a vector of $N$ time-dependent sensitivity signals $\vec{Z}_{k}(t)$
for a single parameter $p_k$ and an $N\times K$-matrix of sensitivity signals when iterating over all parameters.
To ensure full comparability between each sensitivity ${Z}_{i,k}(t)$,
we follow the work of \textcite{kirch_effect_2016} and
propose to normalize the sensitivity with respect to the respective model state and parameter.
According to the definition of the partial derivative,
the ratio between an absolute variation $\Delta x_i$ of the state $x_i(t)$ and
an absolute variation $\Delta p_k$ of the parameter $p_k$ can be approximated through
\begin{equation}\label{linear_approx}
    Z_{i,k}(t) = \frac{\partial x_i(t)}{\partial p_k}\approx \frac{\Delta x_i}{\Delta p_k}.
\end{equation}
We then define the normalized sensitivity as the ratio between a
relative change of the model state $\Delta\tilde{x}_i = \frac{\Delta x_i}{\hat{x}_i}$
and a relative change of the parameter $\Delta\tilde{p}_k = \frac{\Delta p_k}{p_k}$.
$\hat{\vec{x}}$ is the vector of expected maximum values of the states inside the defined ODD.
We hence compute the normalized sensitivity as
\begin{equation}\label{eq:normsens}
	\tilde{Z}_{i,k}(t):=\frac{p_k Z_{i,k}(t)}{\hat{x}_i}
\end{equation}
which fulfills the equation
\begin{equation}\label{eq:normsens_qed}
    \begin{split}
        \tilde{Z}_{i,k}(t) = \frac{p_k Z_{i,k}(t)}{\hat{x}_i} \approx \frac{\Delta x_i}{\Delta p_k} \frac{p_k}{\hat{x}_i}
        = \frac{\Delta x_i}{\hat{x}_i}\frac{p_k}{\Delta p_k} = \frac{\Delta\tilde{x}_i}{\Delta\tilde{p}_i}.
    \end{split}
\end{equation}
Even though the normalization ensures comparability,
it is difficult to make statements about the model's \emph{overall} sensitivity because
this requires to compare the characteristics of each of the $N\cdot K = 10\cdot 49 = 490$
time-dependent sensitivity signals.
We therefore propose to summarize the signals quantitatively
to extract the characteristics of a set of sensitivities.
We argue that the normalization of sensitivities allows their quantitative comparison and hence,
we propose to aggregate the sensitivity signals in each time-point by calculating their sum.
Due to the monotonicity of the summation,
an increase in any single sensitivity contributes to an increase of the determined overall sensitivity.
If it is desired to assess the model sensitivity across all states and parameters at a glance,
the sum over all signals
\begin{equation}\label{eq:sumsens}
	Z_S(t) := \sum_{N}\sum_{K} | \tilde{Z}_{i,k}(t) |
		= \sum_{N}\sum_{K} \frac{p_k | Z_{i,k}(t) |}{\hat{x}_i}
\end{equation}
can be computed.
In this paper, we propose to calculate the sensitivity of a model with respect to selected model states and/or parameters.
This is particularly useful if the sensitivity of specific states and parameters is expected to yield a great impact on the application of the model,
e.g. after prior (sensitivity) analyses have been performed.
Therefore, we summarize the sensitivities for $\tilde{N}\leq N$ selected model states and $\tilde{K}\leq K$ parameters.
We illustrate this approach by shifting the focus of the sensitivity analysis to the vehicle's horizontal motion again:
We select a subset $\vec{\tilde{x}}$ of three model states that directly have an influence in the horizontal plane:
the yaw rate $\psidot(t)$, longitudinal velocity $\vxv(t)$ and lateral velocity $\vyv(t)$.
Based on the specified ODD,
we set $\hat{\vec{x}}=\left (1\,{\mathrm{rad}}/{\mathrm{s}},\, 36.11\,{\mathrm{m}}/{\mathrm{s}},\, 1\,{\mathrm{m}}/{\mathrm{s}}\right )^T$
in \eqref{eq:normsens}.
Furthermore, we focus on the model parameters of the double-track vehicle model
which we assume to be likely to change over the vehicle's lifespan, e.g. due to a variable load or changing road conditions:
the vehicle's mass $m$, yaw moment of inertia $J_z$, distance $l_f$ and the friction coefficient $\mu$.
Hence, in the following section, we evaluate the resulting signal $Z_S(t)$
for a selection of $\tilde{N}\cdot \tilde{K}=3\cdot 4=12$ sensitivity signals.
	\section{Results}
\label{sec:results}
The results of the sensitivity analysis allow to associate the aggregated sensitivity signals $Z_S(t)$
with the ODD and maneuver parameters $m_{\odd}$ that are introduced in the reference trajectory generation.
Due to the large number of combinations for the parameters in $m_{\odd}$,
only selected results are presented in this paper.
We limit the experiments in this section to the lane change maneuver.
As defined in \autoref{tab:parameters} and illustrated in \autoref{fig:reftraj},
we assume a constant lane width $W_{\odd}=4\,\mathrm{m}$ and road curvature $k_{\odd} = 0$ in all examples.

\subsubsection*{Example 1}
\begin{figure}[!h]
    \centering
    \includegraphics[width=\columnwidth]{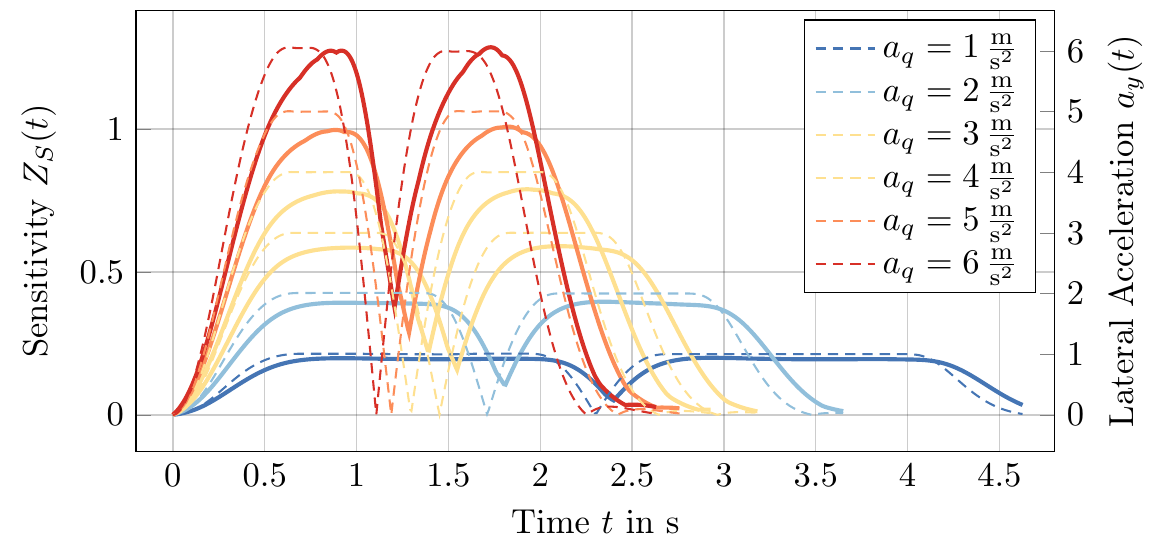}
    \vspace{-0.7cm}
    \caption{Sensitivity $Z_S(t)$ (solid) and lateral acceleration $a_y(t)$ (dashed)
        of the vehicle dynamics model for different acceleration limits $a_{\odd}$
        with $v_{\odd}=50\, \mathrm{km}/\mathrm{h}$ and $\mu_{\odd} = 1$
        during a lane change maneuver}
	\label{fig:lanechange_z_a}
\end{figure}

We first consider the impact of the maximum acceleration $a_{\odd}$ on the model sensitivity.
Furthermore, we set the initial and target speed in the model to
$v_{\odd}:=v_{\odd, 1}=v_{\odd, 2}=13.89\mathrm{m}/\mathrm{s}\approx 50\,\mathrm{km}/\mathrm{h}$ and $\mu_{\odd}=1$.
In \autoref{fig:lanechange_z_a}, the resulting sensitivities $Z_S(t)$ for different values of $a_{\odd}$ are displayed
along with the actual lateral acceleration $a_y(t)$.
We find that the intensity of the lateral acceleration follows the temporal progression of the yaw rate
which, e.g., yields two peaks in case of the lane change maneuver.
As for all examples, we see that the sensitivity follows this pattern as well.
In order to follow the trajectory, we find that the acceleration limit is always completely utilized by the controller.
In particular, due to the constant target speed, the acceleration can be fully applied in the lateral direction since no longitudinal dynamics are required.
The sensitivity measured by $Z_S(t)$ increases monotonously with an increasing acceleration both in terms of its average and maximum value.
When comparing the results to maneuvers where the same available acceleration is fully utilized in the longitudinal direction,
we find that the impact of the longitudinal acceleration is in general negligibly small.

\begin{figure}[!h]
    \centering
    \includegraphics[width=\columnwidth]{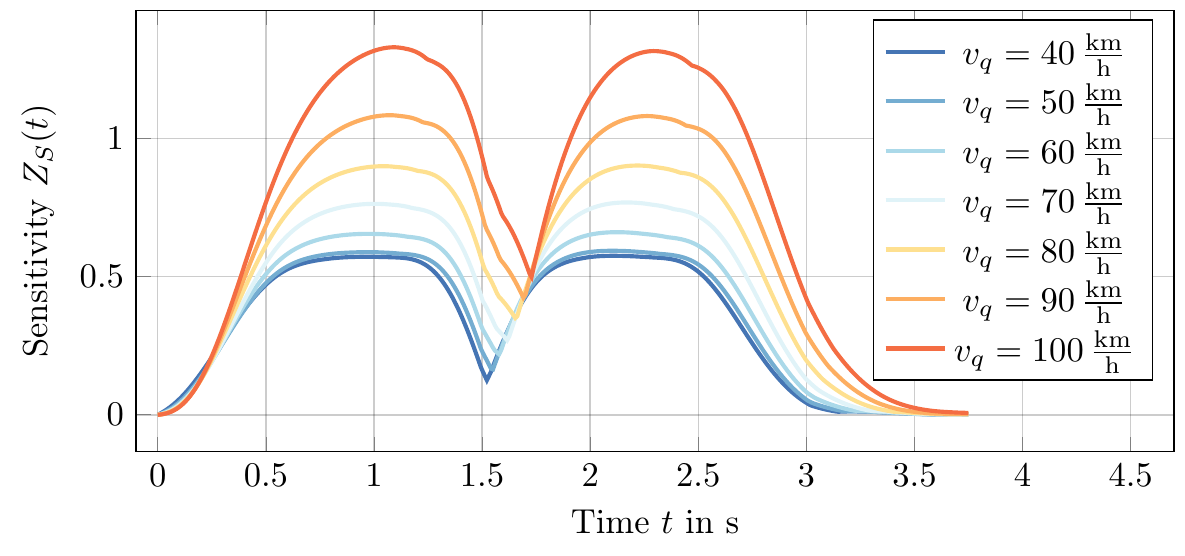}
	\vspace{-0.7cm}
    \caption{Sensitivity $Z_S(t)$ of the double-track vehicle model during a lane change maneuver
        for different constant target speeds $v_{\odd}$
        with fixed acceleration limit $a_{\odd}=3\,\mathrm{m}/\mathrm{s}^2$ and friction coefficient $\mu_{\odd} = 1$}
	\label{fig:lanechange_z_v}
\end{figure}
\begin{figure}[!h]
    \centering

    \includegraphics[width=\columnwidth]{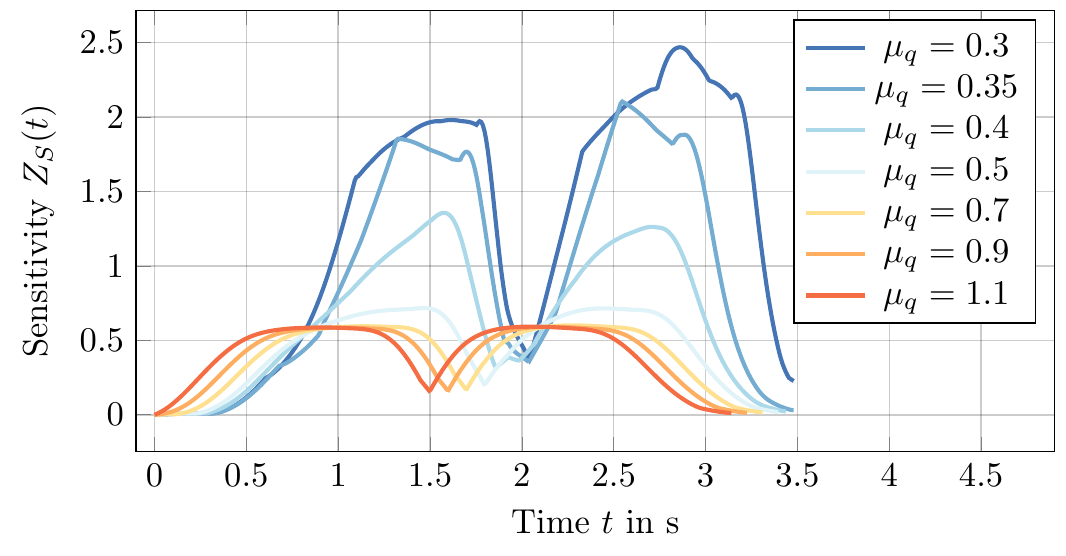}
	\vspace{-0.7cm}
    \caption{Sensitivity $Z_S(t)$ of the double-track vehicle model during a lane change maneuver
        for different friction coefficients $\mu_{\odd}$
        with fixed acceleration limit $a_{\odd}=3\mss$ and speed constraint $v_{\odd}=50\, \mathrm{km}/\mathrm{h}$}
	\label{fig:lanechange_z_mu}
\end{figure}

\subsubsection*{Example 2}
We consider the model sensitivity for different selections of a constant target speed $v_{\odd}=v_{\odd, 1}=v_{\odd, 2}$.
Furthermore, we set $a_{\odd}=3\,\mathrm{m}/\mathrm{s}^2$ and $\mu_{\odd}=1$.
In \autoref{fig:lanechange_z_v}, the resulting sensitivities $Z_S(t)$
for different values of $v_{\odd}$ are displayed with respect to time $t$.
Overall, we find that the sensitivity $Z_S(t)$ increase monotonously with an increasing target speed of $v_{\odd}$
while the acceleration limit is fully utilized in all cases.
Thus, restricting the planned vehicle motion to low speeds can reduce the propagation of model parameter errors into the model behavior.
We particularly find that for large vehicle speeds, the sensitivity signals for a constant acceleration limit of $a_{\odd}=3\mathrm{m}/\mathrm{s}^2$
yield values in a similar order of magnitude compared to the results in \autoref{fig:lanechange_z_a} for a constant vehicle speed
of $v_{\odd} =50\, \mathrm{km}/\mathrm{h}$ but higher acceleration values.
Summarizing example 1 and 2,
we conclude that the model's sensitivity could be lowered at runtime by executing maneuvers with reduced lateral acceleration
when driving at high speeds is required.

\subsubsection*{Example 3}
Finally, we investigate the influence of the friction coefficient $\mu_{\odd}$ on the sensitivity of the model.
The trajectory controller used in this work respects the force limits applied through the road-tire friction
and adapts the resulting control outputs, eventually limiting the vehicle's maximum acceleration.
We set a constant value of $a_{\odd}=3\,\mathrm{m}/\mathrm{s}^2$ and $v_{\odd}=50\,\mathrm{km}/\mathrm{h}$.
While the sensitivity for a fixed acceleration limit and constant speed is almost constant across the higher values of $\mu_{\odd}$,
the model sensitivity increases dramatically for low values $\mu_{\odd}\leq 0.4$ as can be seen in \autoref{fig:lanechange_z_mu}.
As an explanation, we find increasing values in the input functions of the model, e.g. the steering angles $\delta_{i}$,
which also yield abrupt changes occurring in various patterns for different low values of $\mu_{\odd}$.
We thus conclude that the motion controller attempts to compensate for the low friction coefficient and the given maneuver
by applying high control effort, gradually reaching its functional limits.

In all cases, an approximately constant velocity is maintained while the acceleration limit is fully utilized.
In conclusion, this implies that in order to reduce the sensitivity of the model,
neither high acceleration nor high velocity should be applied in the motion controller
when driving in areas which low road-tire friction is required.
However, the sensitivity in \autoref{fig:lanechange_z_mu} reaches much higher values for a low friction coefficient, e.g. $\mu_{\odd}=0.3$,
and a constant acceleration limit and target speed than the results in \autoref{fig:lanechange_z_a} and \autoref{fig:lanechange_z_v}
for a constant friction coefficient $\mu_{\odd}=1$ and higher values of acceleration or speed.
This indicates the need for further analyses within the system design process to investigate the applicability of the model in an ODD
with potentially occurring very low road-tire friction.
	\section{Conclusion and Future Work}
\label{sec:conclusion}
In this paper, we have extended previous work by providing the framework for an ODD-centric sensitivity analysis
which allows considering model sensitivity with respect to selected ODD elements.
As a part of this, we have argued the necessity of considering the vehicle's external behavior
and described the systematical parametrization and instantiation of selected maneuvers.
We have applied the sensitivity analysis framework to a nonlinear double-track vehicle dynamics model
as integrated in a model-predictive trajectory controller.
We have argued for the necessity to normalize the sensitivities for full comparability and
quantitatively summarized them for an aggregated evaluation.
Factors that strongly influence the sensitivity of the model have been quantified in an offline simulation
such as the maximum acceleration, vehicle speed and road-tire friction.
By using the quantitative results of the sensitivity analysis,
sensitivity-minimizing strategies in the context of trajectory and behavior generation can be developed:
While ODD elements such as the weather, road condition and topology, are immutable for the ADS,
the automated vehicle can adapt its behavior, e.g. in terms of its applied target speed or maximum acceleration.
Following the objective to reduce the sensitivity for model-based trajectory controllers and planners,
the sensitivity for a desired maneuver could be estimated
based on the generated a priori knowledge from this simulation framework
and explicitly considered in the generation process, e.g. as a cost or weight to be minimized.
In future work, we therefore aim to conduct further research concerning
the incorporation of model quality assessment in the runtime behavior generation.
	\section*{Acknowledgement}
\label{sec:acknowledgement}
We would like to thank Torben Stolte and Marvin Loba from TU Braunschweig who supported us
through their work on model-predictive trajectory controllers.
Furthermore, we would like to thank Agapius Bou Ghosn from Mines Paris
for the stimulating discussions on vehicle dynamics models.
	%
	\renewcommand*{\bibfont}{\footnotesize} 
	\printbibliography 

@article{andersson_casadi_2019,
 author = {Andersson, Joel AE and Gillis, Joris and Horn, Greg and Rawlings, James B and Diehl, Moritz},
 journal = {Mathematical Programming Computation},
 note = {Publisher: Springer},
 number = {1},
 pages = {1--36},
 title = {{CasADi}: {A} {Software} {Framework} for {Nonlinear} {Optimization} and {Optimal} {Control}},
 volume = {11},
 year = {2019}
}

@inproceedings{bagschik_ontology_2018,
 author = {Bagschik, Gerrit and Menzel, Till and Maurer, Markus},
 booktitle = {{Proc. {IEEE} {Intelligent} {Vehicles} {Symposium} ({IV})}},
 pages = {1813--1820},
 publisher = {IEEE},
 title = {Ontology based scene creation for the development of automated vehicles},
 year = {2018}
}

@misc{bsi_operational_2020,
 author = {{BSI}},
 title = {{PAS} 1883 - {Operational} {Design} {Domain} ({ODD}) {Taxonomy} {For} {An} {Automated} {Driving} {System} ({ADS}) - {Specification}},
 url = {https://www.bsigroup.com/en-GB/CAV/pas-1883/},
 year = {2020}
}

@inproceedings{colwell_automated_2018,
 address = {Changshu},
 author = {Colwell, Ian and Phan, Buu and Saleem, Shahwar and Salay, Rick and Czarnecki, Krzysztof},
 booktitle = {{Proc. {IEEE} {Intelligent} {Vehicles} {Symposium} ({IV})}},
 pages = {1910--1917},
 title = {An {Automated} {Vehicle} {Safety} {Concept} {Based} on {Runtime} {Restriction} of the {Operational} {Design} {Domain}},
 url = {https://ieeexplore.ieee.org/document/8500530/},
 urldate = {2022-03-14},
 year = {2018}
}

@article{dickinson_sensitivity_1976,
 author = {Dickinson, Robert P. and Gelinas, Robert J.},
 journal = {Journal of Computational Physics},
 month = {June},
 number = {2},
 pages = {123--143},
 title = {Sensitivity {Analysis} of {Ordinary} {Differential} {Equation} {Systems} - {A} {Direct} {Method}},
 urldate = {2020-01-16},
 volume = {21},
 year = {1976}
}

@misc{frese_generierung_2019,
 author = {Frese, Matthias},
 language = {Bachelor Thesis, TU Braunschweig},
 title = {Generierung von {Fahrstreifenwechsel}-{Referenztrajektorien} für die {Untersuchung} fehlertoleranter {Trajektorieregelung} in einem urbanen {Umfeld}},
 year = {2019}
}

@misc{gregory_self-aware_2016,
 author = {Gregory, Irene M and Leonard, Charles and Scotti, Stephen J},
 note = {Report},
 title = {Self-aware vehicles: {Mission} and performance adaptation to system health},
 year = {2016}
}

@phdthesis{hamza_contribution_2015,
 author = {Hamza, Sabra},
 month = {July},
 school = {Université de Haute Alsace - Mulhouse},
 title = {Contribution to {Sensitivity} {Analysis} of {Complex} {Systems}: {Application} to {Vehicle} {Dynamics}},
 type = {{PhD} {Thesis}},
 url = {https://tel.archives-ouvertes.fr/tel-01347105},
 year = {2015}
}

@misc{international_organization_for_standardization_iso_2022,
 author = {{International Organization for Standardization}},
 title = {{ISO} 21448 {Road} vehicles - {Safety} of the intended functionality},
 url = {https://www.iso.org/standard/77490.html},
 year = {2022}
}

@article{jang_state_1997,
 author = {Jang, Jin-Hee and Han, Chang-Soo},
 doi = {10.1007/BF02946329},
 issn = {1226-4865},
 journal = {KSME International Journal},
 number = {6},
 pages = {595--604},
 title = {The {State} {Sensitivity} {Analysis} of the {Front} {Wheel} {Steering} {Vehicle}: {In} the {Time} {Domain}},
 volume = {11},
 year = {1997}
}

@misc{jatzkowski_zum_2021,
 author = {Jatzkowski, Inga and Nolte, Marcus and Stolte, Torben and Menzel, Till and Graubohm, Robert and Ernst, Susanne and Steimle, Markus and Salem, Nayel and Richelmann, Jan and Maurer, Markus},
 title = {Zum {Fahrmanöverbegriff} im {Kontext} automatisierter {Straßenfahrzeuge}},
 note = {TU Braunschweig, Report},
 url = {https://www.ifr.ing.tu-bs.de/static/files/forschung/papers/download_pdf.php?id=1156},
 year = {2021}
}

@article{kirch_effect_2016,
 author = {Kirch, Jakob and Thomaseth, Caterina and Jensch, Antje and Radde, Nicole E},
 journal = {EPJ Nonlinear Biomedical Physics},
 note = {Publisher: SpringerOpen},
 number = {1},
 title = {The effect of model rescaling and normalization on sensitivity analysis on an example of a {MAPK} pathway model},
 volume = {4},
 year = {2016}
}

@article{koopman_how_2019,
 author = {Koopman, Philip and Fratrik, Frank},
 journal = {SafeAI @ AAAI},
 title = {How many operational design domains, objects, and events?},
 volume = {4},
 year = {2019}
}

@book{loucks_water_2017,
 author = {Loucks, Daniel P. and van Beek, Eelco},
 isbn = {978-3-319-44234-1},
 publisher = {Springer},
 title = {Water {Resource} {Systems} {Planning} and {Management}},
 year = {2017}
}

@inproceedings{nolte_model_2017,
 author = {Nolte, Marcus and Rose, Marcel and Stolte, Torben and Maurer, Markus},
 booktitle = {{Proc. {IEEE} {Intelligent} {Vehicles} {Symposium} ({IV})}},
 pages = {798--805},
 title = {Model {Predictive} {Control} {Based} {Trajectory} {Generation} for {Autonomous} {Vehicles} – {An} {Architectural} {Approach}},
 year = {2017}
}

@inproceedings{nolte_sensitivity_2020,
 author = {Nolte, Marcus and Schubert, Richard and Reisch, Cordula and Maurer, Markus},
 booktitle = {{Proc. {IEEE} {Intelligent} {Vehicles} {Symposium} ({IV})}},
 pages = {1162--1169},
 publisher = {IEEE},
 title = {Sensitivity {Analysis} for {Vehicle} {Dynamics} {Models} – {An} {Approach} to {Model} {Quality} {Assessment} for {Automated} {Vehicles}},
 year = {2020}
}

@article{nolte_supporting_2020,
 author = {Nolte, Marcus and Jatzkowski, Inga and Ernst, Susanne and Maurer, Markus},
 note = {arXiv pre-print},
 title = {Supporting {Safe} {Decision} {Making} {Through} {Holistic} {System}-{Level} {Representations} \& {Monitoring} - {A} {Summary} and {Taxonomy} of {Self}-{Representation} {Concepts} for {Automated} {Vehicles}},
 url = {https://arxiv.org/abs/2007.13807},
 urldate = {2022-03-23},
 year = {2020}
}

@article{pacejka_magic_1992,
 author = {Pacejka, Hans and Bakker, Egbert},
 doi = {10.1080/00423119208969994},
 journal = {Vehicle System Dynamics},
 note = {Pub.: Taylor \& Francis},
 pages = {1--18},
 title = {The {Magic} {Formula} {Tyre} {Model}},
 volume = {21},
 year = {1992}
}

@inproceedings{polack_guaranteeing_2018,
 address = {Milwaukee},
 author = {Polack, Philip and Altche, Florent and D'Andrea-Novel, Brigitte and de La Fortelle, Arnaud},
 booktitle = {{Proc. {Annual} {American} {Control} {Conference} ({ACC})}},
 doi = {10.23919/ACC.2018.8430886},
 isbn = {978-1-5386-5428-6},
 pages = {3981--3987},
 title = {Guaranteeing {Consistency} in a {Motion} {Planning} and {Control} {Architecture} {Using} a {Kinematic} {Bicycle} {Model}},
 url = {https://ieeexplore.ieee.org/document/8430886/},
 urldate = {2022-04-07},
 year = {2018}
}

@inproceedings{polack_kinematic_2017,
 author = {Polack, Philip and Altché, Florent and d'Andréa-Novel, Brigitte and de La Fortelle, Arnaud},
 booktitle = {{Proc. {IEEE} Intelligent Vehicles Symposium ({IV})}},
 pages = {812--818},
 publisher = {IEEE},
 punctuation = {,},
 title = {The kinematic bicycle model: {A} consistent model for planning feasible trajectories for autonomous vehicles?},
 year = {2017}
}

@incollection{reuter_how_2011,
 author = {Reuter, Hauke and Jopp, Fred and Breckling, Broder and Lange, Christoph and Weigmann, Gerd},
 booktitle = {{Modelling {Complex} {Ecological} {Dynamics}}},
 editor = {Jopp, Fred and Reuter, Hauke and Breckling, Broder},
 isbn = {978-3-642-05028-2 978-3-642-05029-9},
 pages = {323--340},
 publisher = {Springer Berlin Heidelberg},
 title = {How {Valid} {Are} {Model} {Results}? {Assumptions}, {Validity} {Range} and {Documentation}},
 url = {http://link.springer.com/10.1007/978-3-642-05029-9_23},
 urldate = {2020-01-20},
 year = {2011}
}

@misc{sae_j3016_2021,
 author = {{SAE}},
 title = {J3016 - {Taxonomy} and {Definitions} for {Terms} {Related} to {On}-{Road} {Motor} {Vehicle} {Automated} {Driving} {Systems}},
 url = {https://www.sae.org/standards/content/j3016_202104/},
 year = {2021}
}

@article{scholtes_6-layer_2021,
 author = {Scholtes, Maike and Westhofen, Lukas and Turner, Lara Ruth and Lotto, Katrin and Schuldes, Michael and Weber, Hendrik and Wagener, Nicolas and Neurohr, Christian and Bollmann, Martin Herbert and Körtke, Franziska and {others}},
 journal = {IEEE Access},
 pages = {59131--59147},
 title = {6-layer model for a structured description and categorization of urban traffic and environment},
 volume = {9},
 year = {2021}
}

@inproceedings{simkoff_plant-model_2017,
  title={Plant-model mismatch evaluation for unconstrained {MPC} with state estimation},
  author={Simkoff, Jodie M and Wang, Siyun and Baldea, Michael and Chiang, Leo H and Castillo, Ivan and Bindlish, Rahul and Stanley, David B},
  booktitle={Proc. IEEE 56th Annual Conference on Decision and Control (CDC)},
  pages={6177--6182},
  year={2017}
}

@article{stolte_reference_2019,
 author = {Stolte, Torben and Qiu, Lanbin and Maurer, Markus},
 doi = {10.1016/j.ifacol.2019.09.007},
 issn = {24058963},
 journal = {IFAC},
 number = {5},
 pages = {40--47},
 title = {Reference {Trajectories} for {Investigating} {Fault}-{Tolerant} {Trajectory} {Tracking} {Control} {Algorithms} for {Automated} {Vehicles}},
 url = {https://linkinghub.elsevier.com/retrieve/pii/S2405896319306263},
 urldate = {2022-05-04},
 volume = {52},
 year = {2019}
}

@article{stolte_towards_2023,
 author = {Stolte, Torben and Loba, Marvin and Nee, Matthias and Wu, Liren and Maurer, Markus},
 doi = {10.1109/ACCESS.2023.3239518},
 journal = {IEEE Access},
 pages = {1--1},
 title = {Towards {Fault}-{Tolerant} {Vehicle} {Motion} {Control} for {Over}-{Actuated} {Automated} {Vehicles}: {A} {Non}-{Linear} {Model} {Predictive} {Approach}},
 year = {2023}
}

@article{thangavel_handling_2018,
 author = {Thangavel, Sakthi and Subramanian, Sankaranarayanan and Lucia, Sergio and Engell, Sebastian},
 doi = {10.1016/j.ifacol.2018.09.051},
 issn = {24058963},
 journal = {IFAC},
 number = {15},
 pages = {1074--1079},
 title = {Handling {Structural} {Plant}-{Model} {Mismatch} {Using} a {Model}-{Error} {Model} in the {Multi}-{Stage} {NMPC} {Framework}},
 url = {https://linkinghub.elsevier.com/retrieve/pii/S2405896318317105},
 urldate = {2020-01-16},
 volume = {51},
 year = {2018}
}

@phdthesis{tulpule_integrated_2014,
 author = {Tulpule, Punit J},
 school = {Iowa State University},
 title = {Integrated {Robust} {Optimal} {Design} ({IROD}) via {Sensitivity} {Minimization}},
 type = {{PhD} {Thesis}},
 year = {2014}
}

@article{ulbrich_towards_2017,
 author = {Ulbrich, Simon and Reschka, Andreas and Rieken, Jens and Ernst, Susanne and Bagschik, Gerrit and Dierkes, Frank and Nolte, Marcus and Maurer, Markus},
 note = {arXiv pre-print},
 title = {Towards a functional system architecture for automated vehicles},
 year = {2017}
}

@book{werling_neues_2011,
 address = {Karlsruhe},
 author = {Werling, Moritz},
 doi = {10.5445/KSP/1000021738},
 isbn = {978-3-86644-631-1},
 month = {April},
 publisher = {KIT Scientific Publishing},
 title = {Ein neues {Konzept} für die {Trajektoriengenerierung} und -stabilisierung in zeitkritischen {Verkehrsszenarien}},
 year = {2011}
}

@article{yedavalli_controller_1982,
 author = {Yedavalli, Rama Krishna and Skelton, Robert E.},
 journal = {Optimal Control Applications and Methods},
 number = {3},
 pages = {221--240},
 title = {Controller design for parameter sensitivity reduction in linear regulators},
 url = {https://onlinelibrary.wiley.com/doi/10.1002/oca.4660030302},
 urldate = {2022-04-11},
 volume = {3},
 year = {1982}
}
\end{document}